\documentclass[a4paper,11pt]{article}
\usepackage{pos}

\title{Constraining models of hadronic showers using proton-Oxygen collisions at the LHC involving proton/neutron tagging}
 \ShortTitle{Constraining models of hadronic showers at the LHC}

\author*[a,b]{Michael Pitt}

\affiliation[a]{Ben-Gurion University of the Negev, Department of Physics,
  Beer-Sheva, Israel}

\affiliation[b]{The University of Kansas, Department of Physics,
Lawrence, USA}

\emailAdd{michael.pitt@cern.ch}

\abstract{
The study of hadronic showers, which are produced by cosmic rays penetrating the Earth's atmosphere, is essential for shedding light on the origins and characteristics of high-energy particles originating from space and reaching our planet. At the Large Hadron Collider at CERN, there are plans to conduct a short run of proton--oxygen collisions in 2025 to refine the modeling of hadronic showers. This work explores the potential impact on constraining models of hadronic showers by measuring interactions facilitated by color-neutral objects such as photons, pomerons, and pions. These interactions are often characterized by high-energy protons or neutrons produced at forward rapidities and can be tagged using dedicated forward proton and neutron detectors.
}

\FullConference{%
XVIII International Conference on Topics in Astroparticle and Underground Physics (TAUP2023)\\
 28.08-01.09.2023\\
University of Vienna\\}

\begin{document}
\maketitle

\section{Introduction}

\begin{figure}[htb]
\centering
\includegraphics[width=0.40\textwidth]{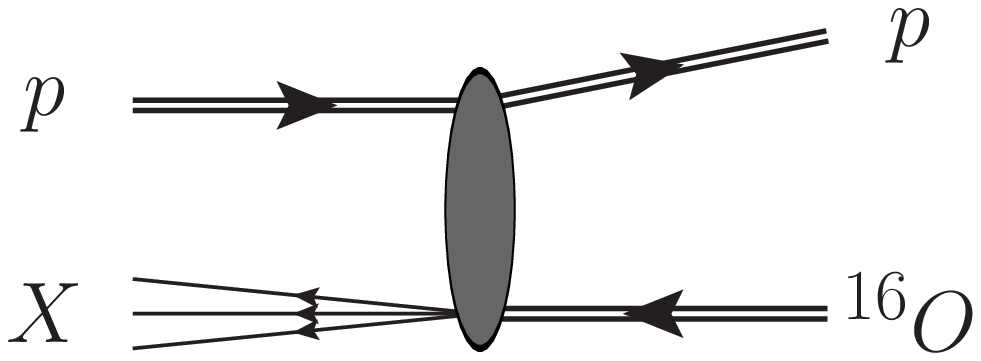}
\hspace{0.5cm}
\includegraphics[width=0.40\textwidth]{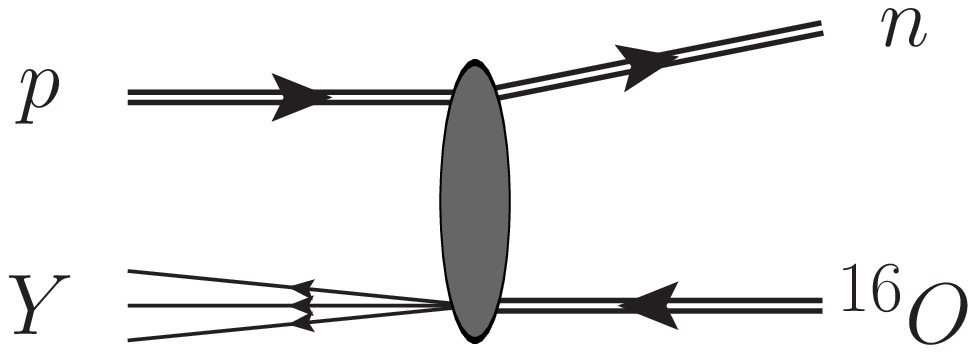}
\caption{Schematic diagrams of $pO$ collisions with an intact proton (left) or a neutron (right) produced at very forward rapidities.\label{fit:process}}
\end{figure}

A vital component in studying the nature of Cosmic Rays (CRs) is the determination of the mass and energy spectra by measuring profiles of the air showers produced by the CRs and matching them to profiles predicted by hadronic Monte Carlo (MC) simulations, which are often tuned using data from the Large Hadron Collider (LHC) \cite{mc1,mc2,mc3,mc4}. Nonetheless, discrepancies between different model predictions persist, even at LHC energies, leading to large uncertainties in understanding the composition of CRs \cite{Supanitsky:2022zcw}. A short run of proton--oxygen ($pO$) collisions is scheduled during LHC Run 3 \cite{Bruce:2021hjk}, aiming to improve the modeling of hadronic interactions. The primary focus of the standard research program at the LHC will be on non-diffractive interactions \cite{Brewer:2021kiv}; however, diffractive interactions, or in general any interaction involving color-neutral objects, can also be explored by tagging forward protons in $pO\to pX$ interactions or forward neutrons in $pO\to nX$ interactions (where X represent the dissociation products for the oxygen ion), providing a unique opportunity to study those components with better precision. Figure~\ref{fit:process} illustrates schematic diagrams of processes of interest.

The color-neutral interactions are weakly constrained at the LHC, resulting in substantial discrepancies between the experimental data and the predictions of MC simulations (e.g.,  in proton-lead collisions \cite{CMS:2023lfr}), suggesting there are missing interactions not included in these event generators, which can be further probed using forward neutron and proton detectors.

\section{Forward proton and neutron detectors at the LHC} 

The ATLAS forward proton (AFP) detector \cite{Adamczyk:2015cjy} and the CMS-TOTEM Precision Proton Spectrometer (CT-PPS) \cite{CMS:2014sdw}, deployed during LHC Run 2 (2015--2018) by the ATLAS and the CMS collaborations, are specialized near-beam detectors located approximately 200 meters from the interaction point (IP). These detectors were operational throughout the standard high-luminosity runs at the LHC and delivered a broad range of physics results primarily dedicated to studying the central exclusive production processes in proton--proton collisions \cite{Sopczak:2023xyd,Royon:2023ihe}.

The kinematic acceptance for forward protons is contingent upon the arrangement of the LHC magnetic field. During  standard LHC runs, protons that lose  1.5\%--15\% of their momentum are consequently deflected from their trajectories into the proton detectors. A similar arrangement is expected to apply to the upcoming $pO$ run, providing the capability to measure the diffractive component of the total $pO$ cross-section by tagging the intact protons.

The zero-degree calorimeter (ZDC) is a specialized detector positioned at a zero-degree angle relative to the beamline,  substantially contributing to the heavy-ion physics program at the LHC~\cite{Suranyi:2021ssd,ATLAS:2017fur}. Its primary function is to detect the forward neutral particles produced in the collisions, primarily spectators from ion disintegration. The ZDC detectors for both ATLAS and CMS experiments are hosted in a dedicated slot inside the neutral beam absorbers (TAN) at a distance of 140~meters from the IP, which shields the LHC machine components against neutral particles emerging from the IP. The ZDC is capable of detecting forward neutrons and photons with pseudorapidities greater than 8.5, and it consists of an electromagnetic section, approximately 30 radiation lengths long, and three hadronic modules, each about 1.15 interaction lengths long \cite{Suranyi:2021ssd}. 

\section{New constrains on MC hadronic models}

\begin{figure}[htb!]
\centering
\includegraphics[width=0.4\textwidth]{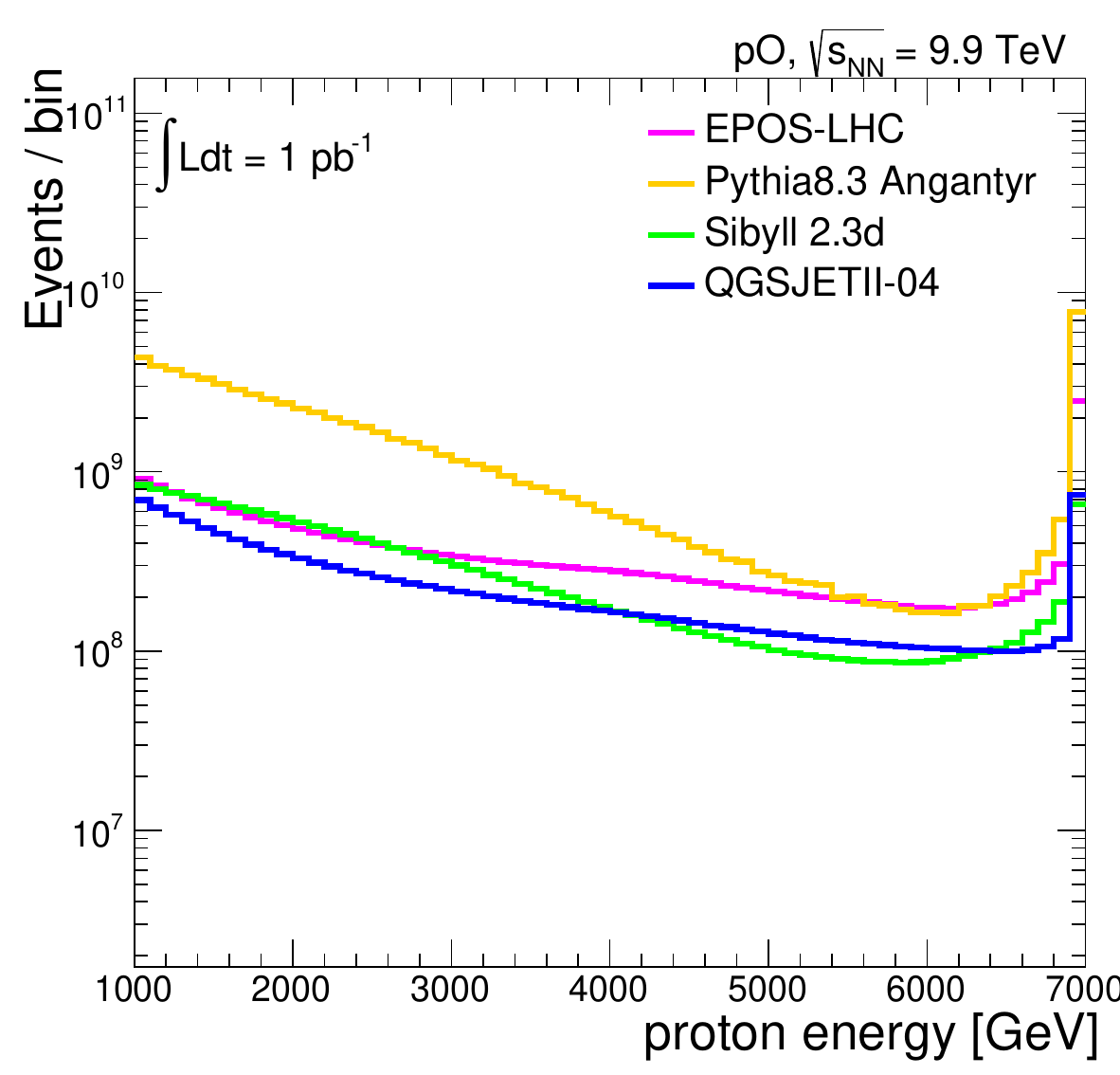} 
\includegraphics[width=0.4\textwidth]{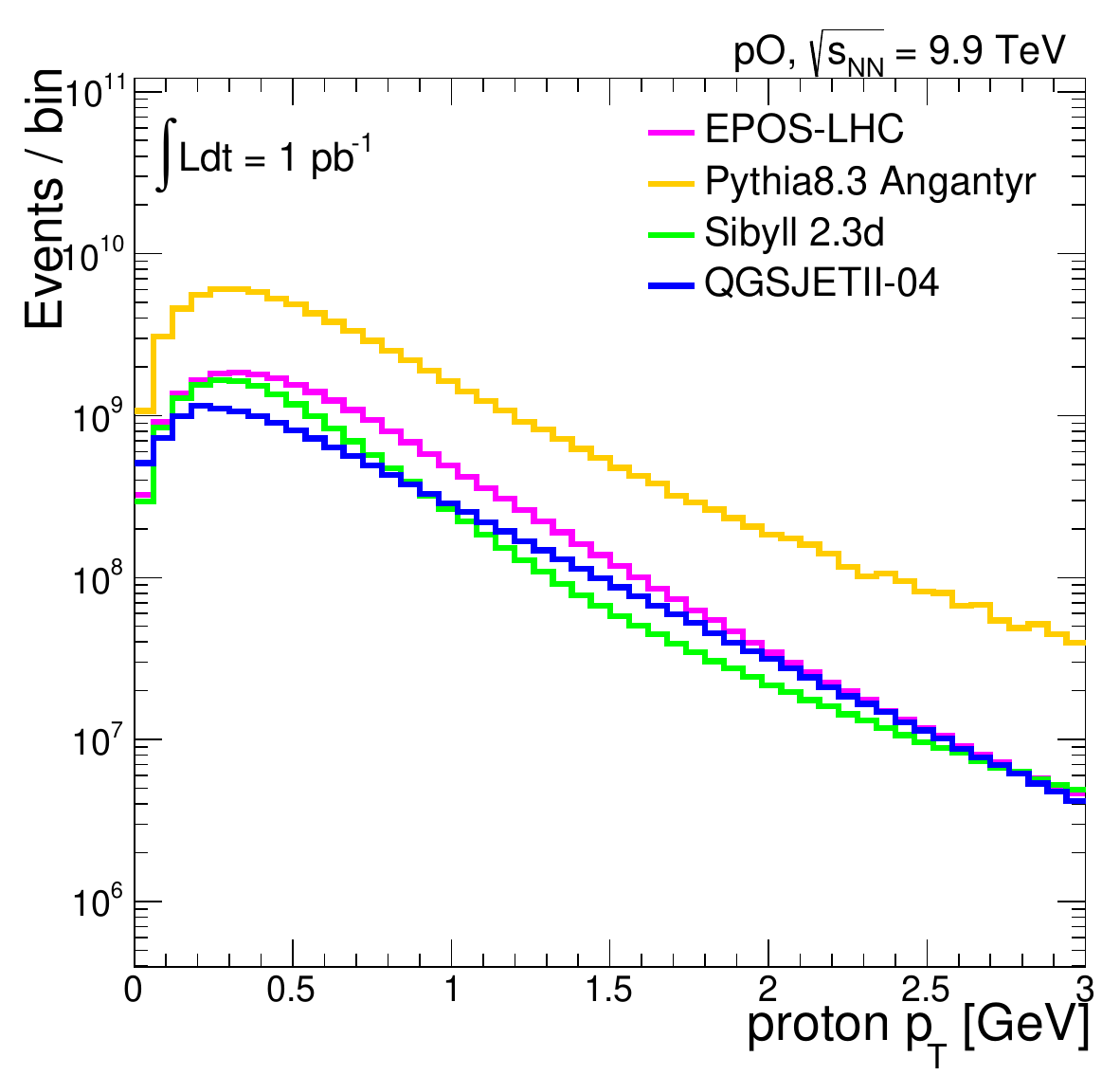}\\
\includegraphics[width=0.4\textwidth]{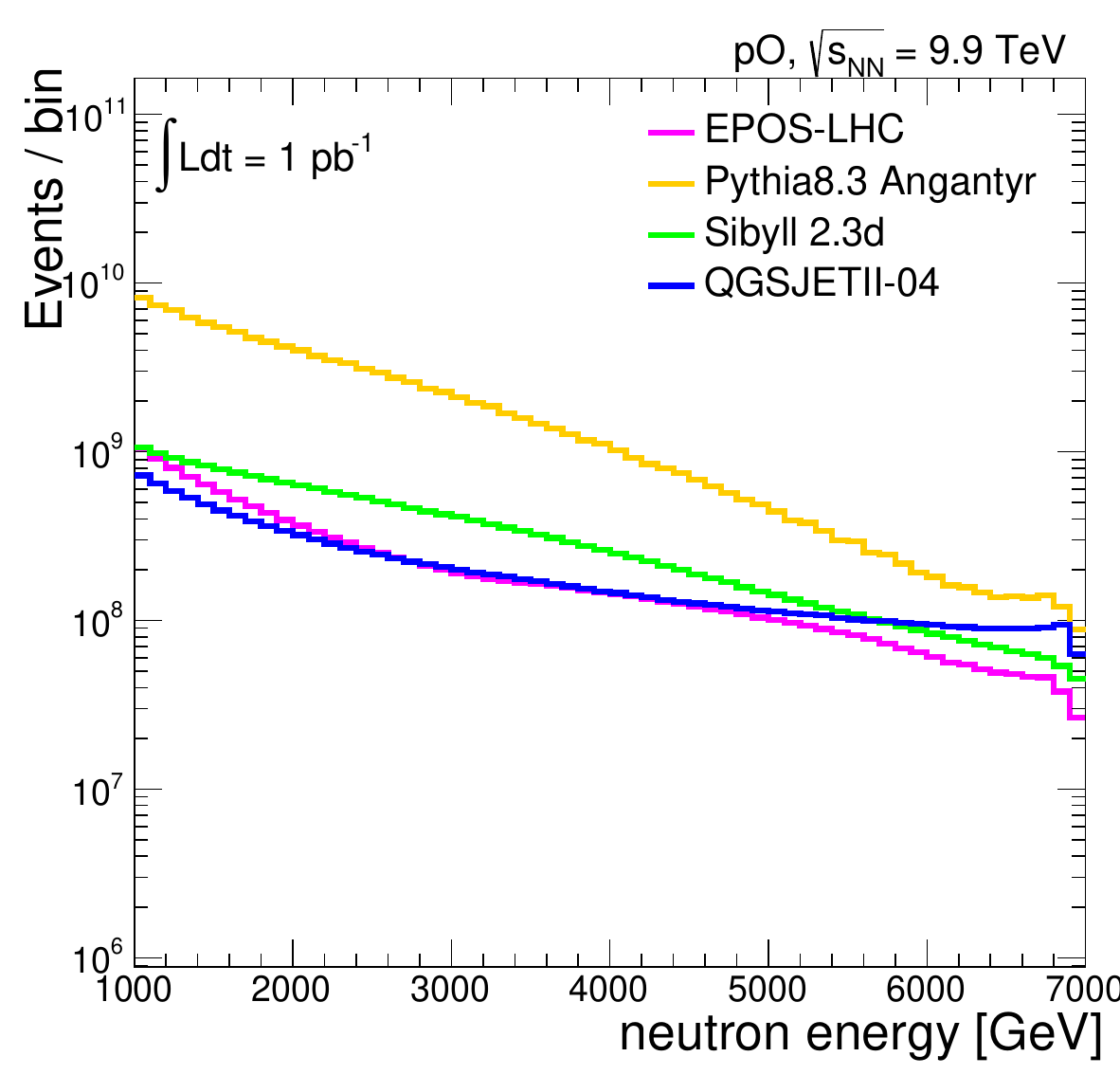} 
\includegraphics[width=0.4\textwidth]{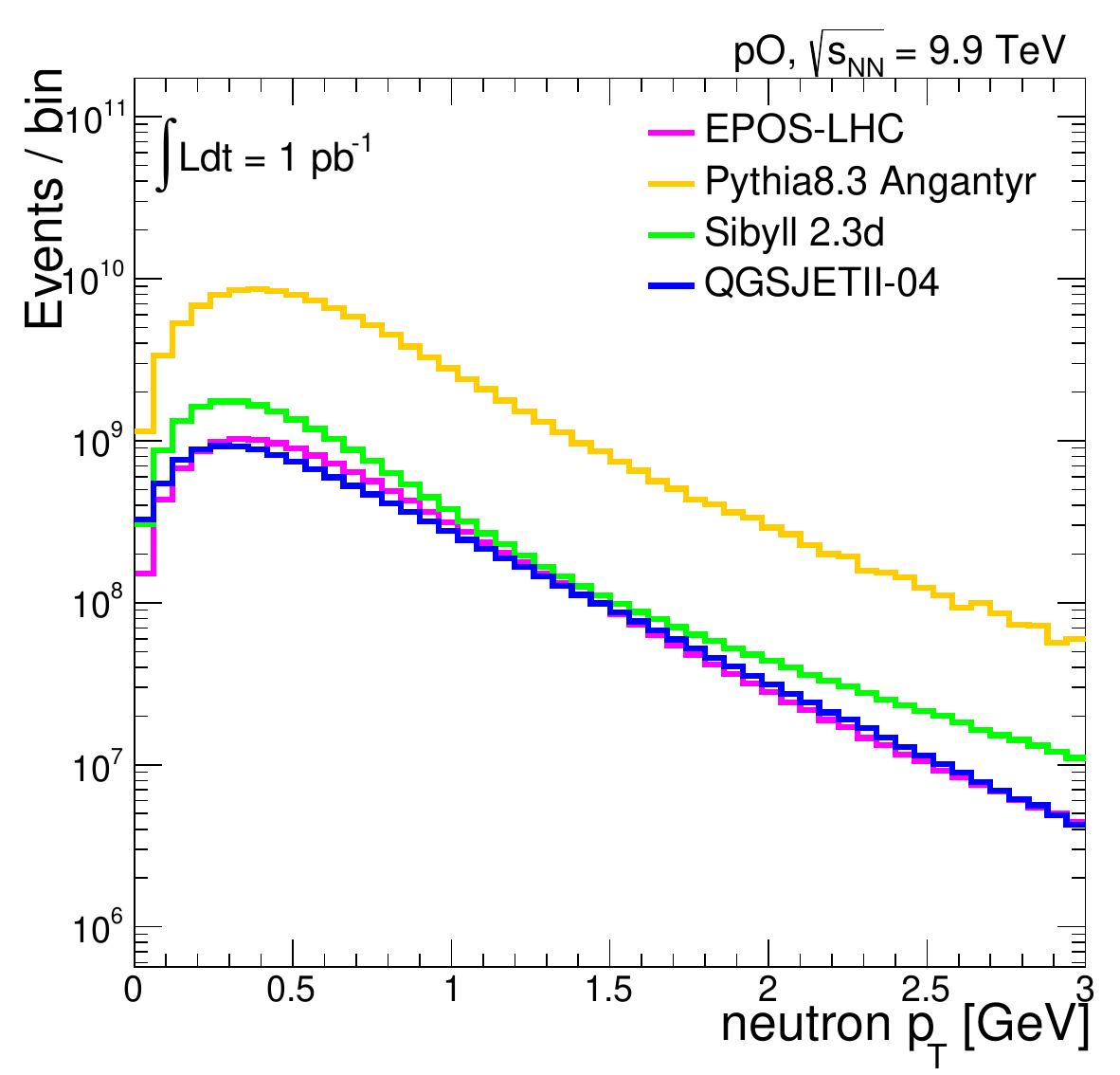}
\caption{Differential cross-section as a function of nucleon energy (left) and transverse momentum (right) for protons (top) and neutrons (bottom), obtained using the EPOS-LHC (magenta), Pythia8 Angantyr (yellow), Sibyll 2.3d (green), and QGSJETII-04 (blue) MC event generators.\label{Fig:forward_kinematics}}
\end{figure}

Colorless interactions (including elastic, diffractive, and pion exchange processes), accounting for approximately 20\% of the total $pO$ cross-section, were simulated using different MC event generators at $\sqrt{S_\text{NN}}=9.9$~TeV assuming an integrated luminosity of $L_\text{int}=1nb^{-1}$ as it is expected during the $pO$ run in 2025.  The typical spectra of forward protons and neutrons are depicted in Figure~\ref{Fig:forward_kinematics}, suggesting a high event rate within the kinematic range where the discrepancy between the models is observed.

\begin{figure}[htb!]
\centering
\includegraphics[width=0.6\textwidth]{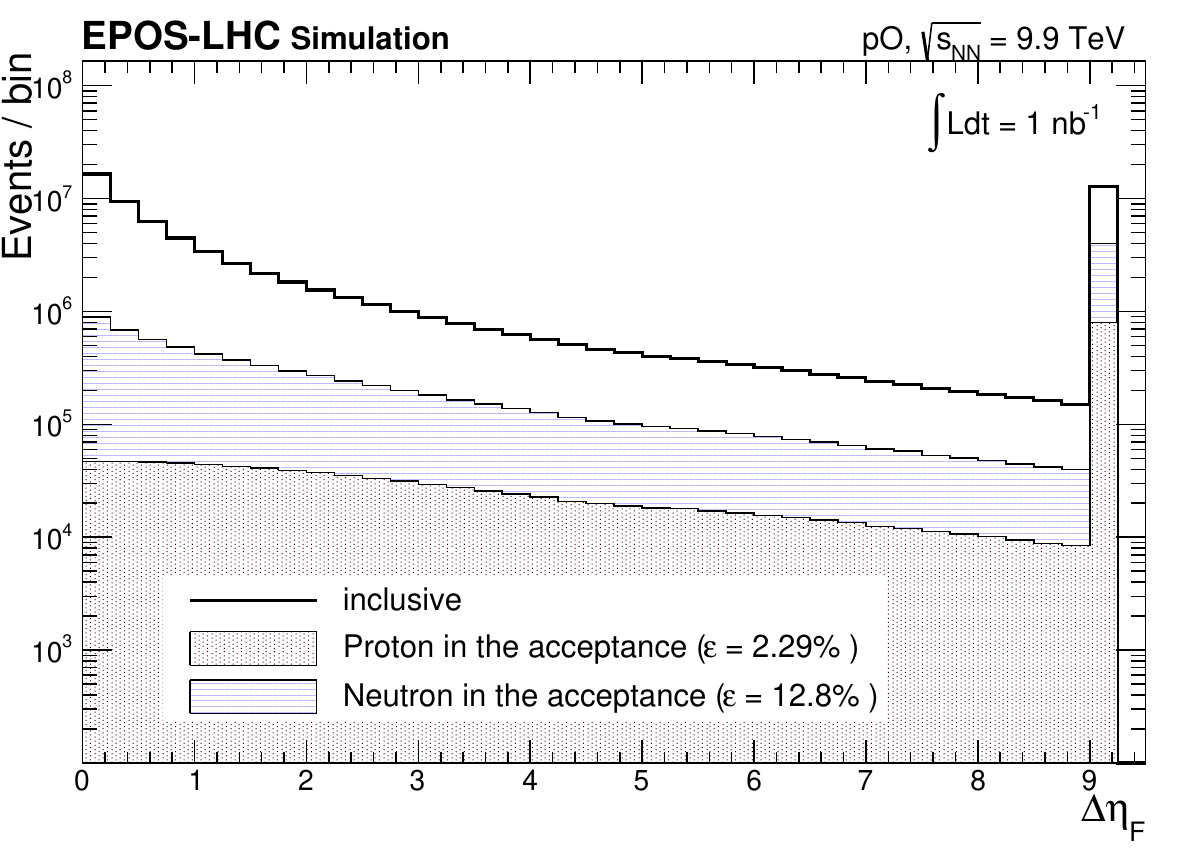}
\caption{Contribution from $pO\to pX$ and $pO\to nX$ interactions with a proton or neutron within the detector acceptance. The peak in the $\Delta\eta_\text{F}$ distribution corresponds to zero-bias events (i.e., no particles with an energy above 1 GeV are detected within the pseudorapidity range of $|\Delta\eta| < 4.5$). Figure adopted from ref.~\cite{Pitt:2023qmd}.\label{Fig:fig3}}
\end{figure}

Interactions involving color-neutral mediators are also characterized by large gaps in the rapidity distribution of the final-state particles, denoted as $\Delta\eta_F$. In contrast to non-diffractive inelastic events, where the probability of finding a continuous rapidity region $\Delta\eta_F$ that is free of particles is exponentially suppressed, color-neutral interactions such as those involving pomeron or pion exchange are distinctive for their apparent rapidity gaps, and discriminating between these topologies (pomeron or pion exchange) is only achievable through proton and neutron tagging. Figure~\ref{Fig:fig3} illustrates the contribution from processes, with a proton or a neutron in the acceptance, as a function of  $\Delta\eta_\text{F}$.

\section{Conclusions}

Forward neutron and proton detectors are expected to become increasingly vital during  upcoming oxygen collisions, augmenting the scope of the existing physics research program. These detectors will extend the measured phase space range, which is essential for refining our understanding of color-neutral interactions. They are poised to provide precise constraints on diffractive and elastic interactions in proton--ion collisions, including the first measurement of the elastic component of proton--oxygen interactions.

\end{document}